\documentclass{PoS}

\usepackage{natbib} 
\usepackage{graphicx}
\usepackage{txfonts}
\usepackage{epsfig}
\def\inte{{\em INTEGRAL}}

\def\rxte{{\em RXTE}}
\def\swift{{\em Swift}}

\usepackage{amssymb}
\usepackage{natbib}
\usepackage{aas_macros} 
\usepackage{threeparttable}

\title{Long-term soft X-ray characterization of Supergiant Fast X-ray Transients: the cumulative luminosity distributions}

\ShortTitle{Long-term soft X-ray characterization of Supergiant Fast X-ray Transients}

\author{\speaker{Enrico Bozzo}\\
        ISDC, University of Geneva, Chemin d\'Ecogia 16, CH-1290 Versoix, Switzerland
        E-mail: \email{enrico.bozzo@unige.ch}}

\author{Patrizia Romano\\
        INAF, Istituto di Astrofisica Spaziale e Fisica Cosmica - Palermo, via U. La Malfa 153, 90146 Palermo, Italy\\
        E-mail: \email{romano@ifc.inaf.it}}
        
 \author{Lorenzo Ducci\\
        Institut f\"ur Astronomie und Astrophysik, Eberhard Karls Universit\"at, Sand 1, 72076 T\"ubingen, Germany\\ 
        ISDC, University of Geneva, Chemin d\'Ecogia 16, CH-1290 Versoix, Switzerland\\
        E-mail: \email{Lorenzo.Ducci@unige.ch}}       
        
 \author{Federico Bernardini\\
        NYUAD, PO Box 129188, Abu Dhabi, United Arab Emirates\\
        INAF, Osservatorio Astronomico di Capodimonte, Salita Moiariello 16, I-80131 Napoli, Italia\
        E-mail: \email{fb61@nyu.edu}}  
             
\author{Maurizio Falanga\\
        International Space Science Institute, Hallerstrasse 6, CH-3012 Bern, Switzerland; 
International Space Science Institute in Beijing, No.~1 Nan Er Tiao, Zhong Guan Cun, Beijing 100190, China.\\
        E-mail: \email{mfalanga@issibern.ch}}

\abstract{We constructed the cumulative luminosity distributions of most supergiant fast X-ray transients (SFXTs) 
  and the classical supergiant X-ray binary (SgXB) IGR\,J18027-2016 by taking advantage of the long term monitoring of these sources 
  carried out with \swift\,/XRT (0.3-10~keV). Classical SgXBs are characterized by cumulative 
  distributions with a single knee around $\sim$10$^{36}$-10$^{37}$~erg~s$^{-1}$, while all SFXTs are found to be significantly 
  sub-luminous and the main knee in their distributions is shifted at lower luminosities 
  ($\lesssim$10$^{35}$~erg~s$^{-1}$). As the latter are below the sensitivity limit of large field 
  of view instruments opearating in the hard X-ray domain ($>$15~keV), we show that a soft X-ray monitoring 
  is required to reconstruct the entire profile of the SFXT cumulative luminosity distributions.   
  The difference between the cumulative luminosity distributions of classical SgXBs and SFXTs 
  is interpreted in terms of different wind accretion modes.}

\FullConference{10th INTEGRAL Workshop: "A Synergistic View of the High Energy Sky" - Integral2014,\\
		15-19 September 2014\\
		Annapolis, MD, USA }

\begin{document}

\section{Introduction}
\label{sec:intro}
  
Most of the so-called ``classical'' Supergiant X-ray binaries (SgXBs) host a neutron star (NS) accreting material from the wind of its 
O-B supergiant companion. These sources are characterized by a nearly persistent 
X-ray luminosity of $L_{\rm X}$=10$^{35}$-10$^{37}$~erg~s$^{-1}$ (mostly depending on their orbital period) and display variations in the 
X-ray intensity by as large as a factor of $\sim$20-50 on time scales of hundreds to thousands of seconds. This pronounced 
variability is usually ascribed to the presence of inhomogeneities in the accreting medium (``clumps''). 

Supergiant Fast X-ray transients (SFXTs) are a sub-class of SgXBs, displaying a much more pronounced variability in the 
X-ray domain: they spend most of their time in low luminosity states ($L_{\rm X}$=10$^{32}$-10$^{33}$~erg~s$^{-1}$) and only sporadically 
undergo a few-hours long outbursts reaching peak luminosities comparable to the persistent level of other SgXBs \citep{sguera06}. 
As inhomogeneities in the accreting material are not sufficient to account for such pronounced variability, the mechanism regulating 
the SFXT activity is still a matter of debate \citep[see, e.g.,][]{bozzo13}. 

Large field-of-view (FoV) hard X-ray imagers, like the IBIS/ISGRI on-board \inte\ \citep[20~keV-1~MeV;][]{ubertini03,lebrun03} and \swift\,/BAT 
\citep[15-150~keV][]{barthelmy05}, have been very efficient in catching a large number of sporadic SFXT outbursts. The long-term monitoring data 
now available have been exploited to estimate the SFXT activity duty-cycle \citep[see, e.g.,][hereafter P14]{romano14,paizis14}. 
The latter was found to be significantly lower (1-5~\%) in the hard X-ray domain than that of classical SgXBs ($\sim$80~\%). 
By using all archival ISGRI data, P14 also reported a detailed comparison between the cumulative 
luminosity distributions of these two classes of sources. They showed that in the energy range 17-50~keV the distributions of SFXTs can 
be reasonably well described by a single power-law, while those of classical SgXBs are typically more complex, showing a knee at 
luminosities $\sim$10$^{36}$~erg~s$^{-1}$ and requiring at least two different power-laws to satisfactorily describe their profiles.    

The fainter states of SFXTs can be studied within a reasonable sensitivity only by using pointed observations with focusing X-ray telescopes. 
Among these, XRT \citep{burrows05} on-board \swift\ proved to be particularly useful in carrying out long-term 
monitoring of the SFXTs, as it can take advantage of the unique scheduling flexibility of the \swift\ satellite \citep{gehrels05}. 
For most of the SFXTs, bi-weekly observations lasting 1~ks and achieving a limiting sensitivity comparable to their lowest emission 
level have been carried out from 2007 to present \citep{sidoli08,romano09,romano11}. These data provide now a sufficiently long baseline to 
be compared with the results obtained through wide FoV hard X-ray imagers.  
A first comparison was reported by \citet[][hereafter R14]{romano14b}, who showed that XRT data allow us to estimate the SFXT  
duty cycle across 4 orders of magnitude in X-ray luminosity. In this letter, we concentrate on their soft X-ray cumulative 
luminosity distributions.

\section{\swift\ sample and data analysis}
\label{sec:data}

We made use of all available XRT data collected from 2007 to 2013 from the 10 SFXTs reported below. 
This data-set has been presented in \citet{bozzo14} and comprises:  \\
- data from the monitoring of the SFXTs IGR\,J16479-4514, XTE\,J1739-302, and IGR\,J17544-2619  
collected from 2007-10-26 to 2009-11-03;  \\
- data from the monitoring of AX\,J1841.0-0536 collected from 2007-10-26 to 2008-11-15; \\
- data from the monitoring of one complete orbit of the SFXTs IGR\,J18483-0311 (carried out from 2009-06-11 to 2009-07-08), 
IGR\,J16418-4532 (carried out from 2011-02-18 to 2011-07-30), and IGR\,J17354-3255 (carried out from 2012-07-18 to 2012-07-28); \\
- data accumulated during the most recent monitoring campaigns of the SFXTs IGR\,J08408-4503, IGR\,J16328-4726, and IGR\,J16465-4507. 
These campaigns have been carried out from 2011-10-20 to 2013-10-24. 
  
In order to compare results from the SFXTs to those of classical neutron star SgXBs, we also analyzed XRT data obtained 
during a monitoring campaign of IGR\,J18027-2016 \citep[for a detailed log of the data used, see][]{bozzo14}. 
This is a classical eclipsing SgXBs with an orbital period of 4.57~days. The neutron star spin period is 
139.47~s \citep{hill05}, and the companion star was classified as a B0-BI supergiant 
located at a distance of 12.4~kpc \citep{mason11}. 
The XRT monitoring campaign covered 7 orbital periods of the source with daily pointings 
of 1--2\,ks in Photon counting (PC) mode. These observations also provided
the most accurate source position to date for this source 
at  RA(J$2000)=270.67494$,  Dec(J$2000)= -20.28813$ (estimated uncertainty 1.4'' radius).  
We extracted from the XRT data the 0.3--10\,keV background-subtracted lightcurve 
(100~s resolution) and the average spectrum. 
The latter could be fit ($\chi^2_{\rm red}$/d.o.f. = 0.97/98) by using an absorbed power-law model with a 
column density of $(2.6\pm0.2)\times10^{22}$~cm$^{-2}$ 
and a photon index of $\Gamma$ = 0.43$\pm$0.09. An emission Fe K$\alpha$ line was also added to the fit. 
The estimated centroid energy of the line is 6.39$\pm$0.06~keV and the corresponding equivalent width $\sim$0.1~keV.

\section{XRT cumulative luminosity distributions}
\label{sec:distributions}

\begin{figure*}
\centering
\includegraphics[scale=0.4,angle=90]{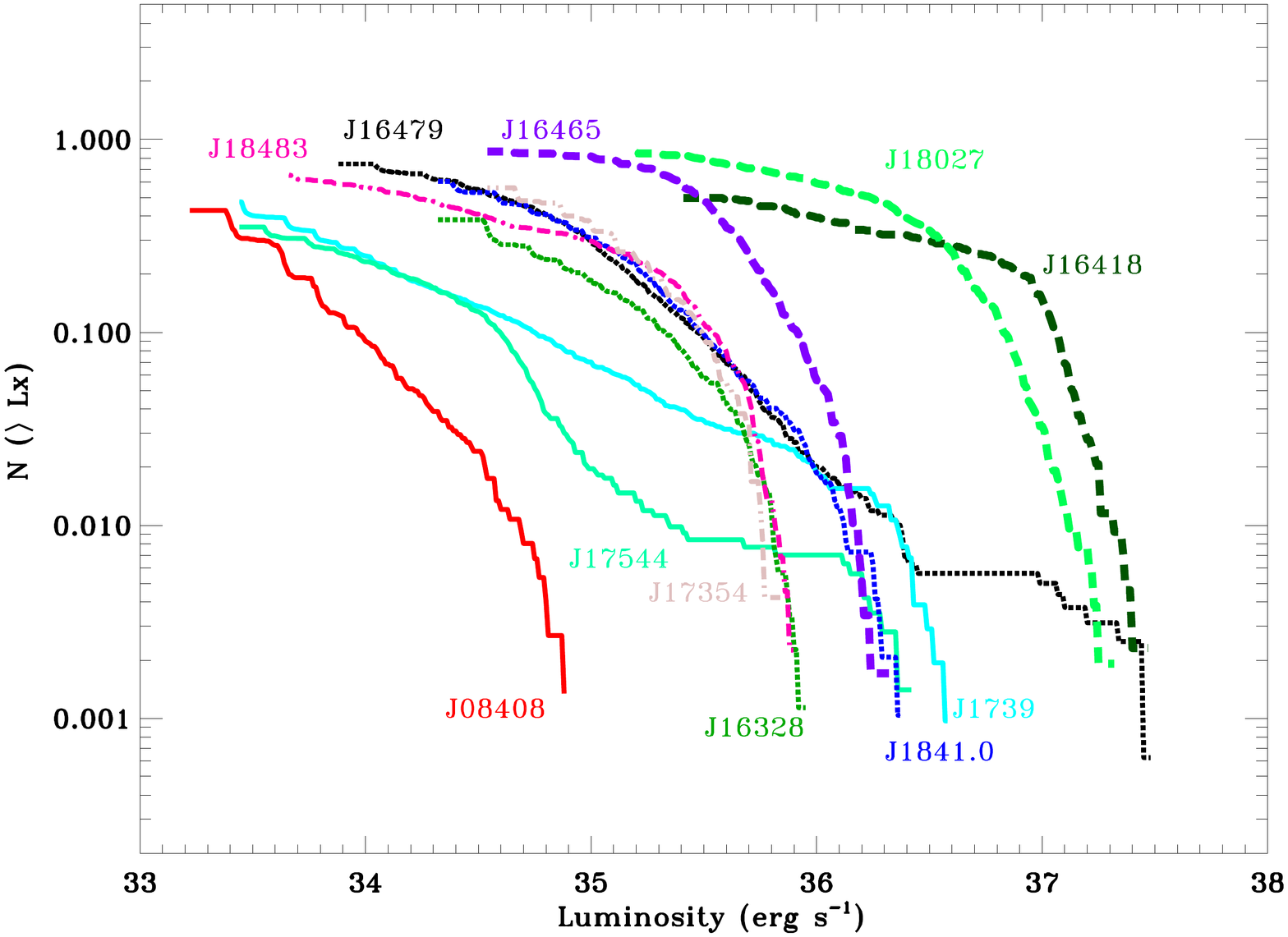}
\includegraphics[scale=0.4,angle=90]{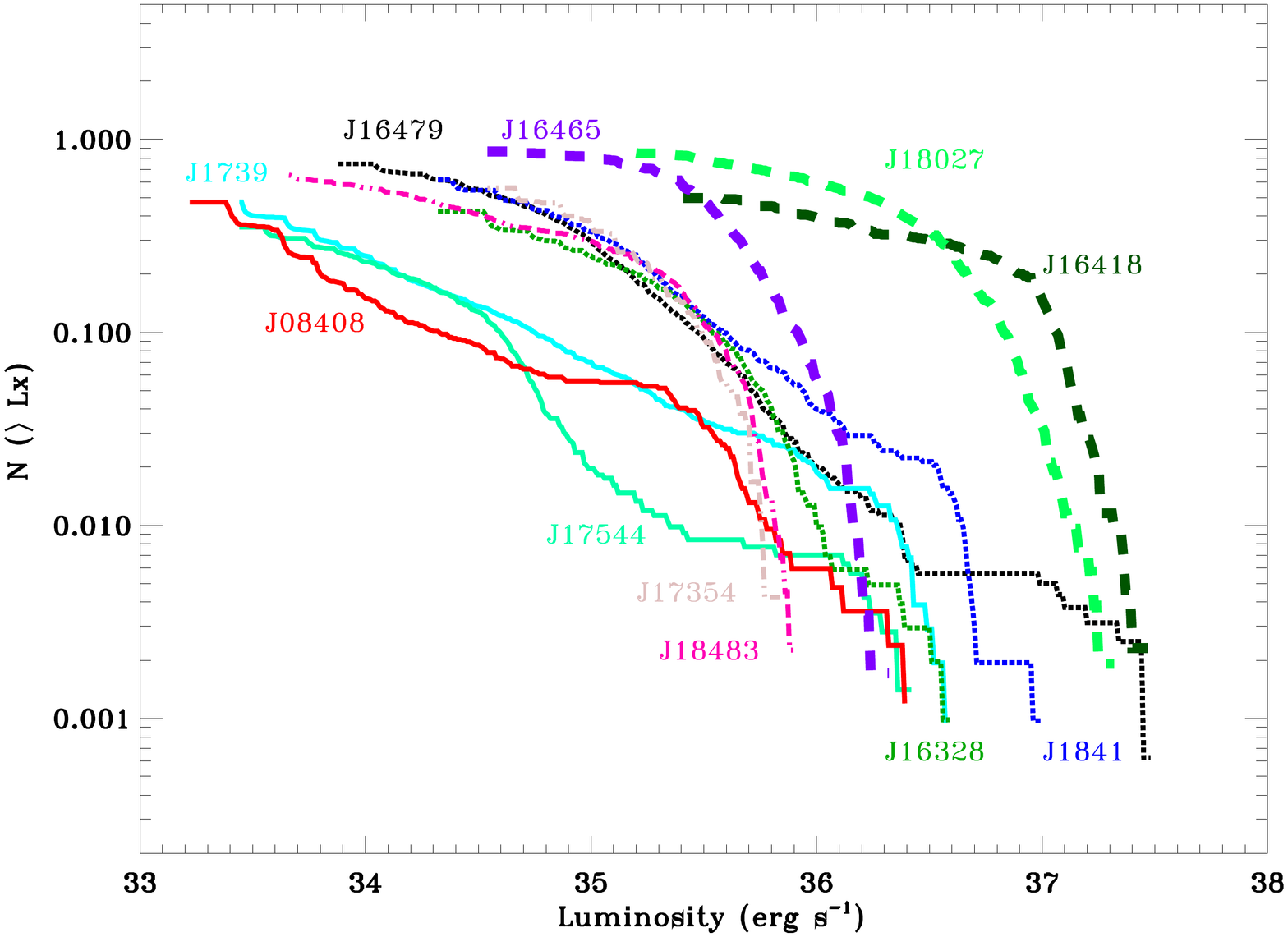}
\caption{Top: cumulative luminosity distributions of all sources considered in this work. 
The distributions are constructed in the 2-10~keV energy band but using only XRT data collected during the 
monitoring campaigns of all sources. Bottom: same as for the figure on the top, but in this case we also considered for the sources 
IGR~J08408$-$4503,  IGR~J16328$-$4726, and AX~J1841.0$-$0536 the outbursts recorded by XRT outside the corresponding monitoring campaigns. 
In both cases we represented the cumulative luminosity distributions of classical SgXBs with thicker dashed lines (including 
IGR~J16418$-$4532 and IGR~J16465$-$4507, and used dot-dashed lines for the intermediate SFXTs. 
The distributions of SFXTs have been represented with dotted lines, while solid lines have been used for the three most extreme SFXTs 
IGR~J08408$-$4503, XTE\,J1739-302, and IGR\,J17544-2619.}      
\label{fig:cumul} 
\end{figure*}

We created the cumulative luminosity distributions of all sources considered in this work by using 
the corresponding XRT lightcurves binned at 100 s. Observations where a significant detection of the source 
($\geq$3~$\sigma$) was not achieved in 100 s were excluded from further analysis (including time intervals corresponding 
to X-ray eclipses, where relevant). 
For all SFXTs, we used the same distances as R14 to convert from count-rates to luminosity and   
the 2-10~keV unabsorbed flux of each source. The conversion for IGR J18027-2016 was calculated by adopting  
the parameters obtained from the fit to the mean source spectrum.   
The cumulative luminosity distributions of all SFXTs that have been monitored 
at least for one orbital period by XRT and that of IGR J18027-2016 
are shown in Fig.~\ref{fig:cumul} with 100 bins per decade in luminosity. 
We need to distinguish the following cases: (i)  
IGR J16479$-$4514, XTE J1739$-$302, and IGR J17544$-$2619 went into 
outburst during the corresponding observing campaigns, so the data shown in the top panel of 
Fig.~\ref{fig:cumul} include all luminosity levels experienced by these sources; (2)  
IGR J08408$-$4503,  IGR J16328$-$4726, and AX J1841.0$-$0536
did not experience an outburst during the monitoring, but outbursts were recorded 
at different times (R14). To assess their overall 
distributions, we thus also added the data of such outbursts and plotted the corresponding distributions in 
the bottom panel of Fig.~\ref{fig:cumul}. This does not affect our conclusions. 

All cumulative luminosity distributions in Fig.~\ref{fig:cumul} were also normalized to the total exposure time of each source, such that the 
source DC correspond to the highest value on the y-axis and an easier comparison can be carried out with the cumulative 
distributions obtained in the hard X-rays (P14).  
By comparing our Fig.~\ref{fig:cumul} with Fig.~1 in P14, we first notice that the cumulative distributions of SFXTs 
in the soft X-rays do not have power-law shaped profiles. More precisely: 
\begin{itemize}
\item the source IGR J18027-2016 is characterized by a cumulative distribution with a single knee around 10$^{36}$ erg s$^{-1}$, as expected 
for classical SgXBs. 

\item the distributions of IGR J16465-4507 and IGR J16418-4532 closely resemble 
those of classical SgXBs. In the logN-logL plot, a knee is observed at a certain critical luminosity and the slope of the profile 
changes abruptly above this value. Similar profiles were observed by P14 in the cases of Vela X-1 and 
4U 1700-377. The only difference seems to be that the IGRs mentioned above are at a distance much larger than that of Vela X-1 and 
4U 1700-377. Their fluxes are thus too low (by a factor of $\gtrsim$100) for the wide FoV instruments to exploit their entire X-ray dynamical 
range. The higher sensitivity of XRT allows us to follow more accurately their activity and the 
complete profile is recovered. Interestingly, IGR J16465-4507 and IGR J16418-4532 have been recently 
classified as classical SgXBs rather than SFXTs by R14 and \citet{drave13}, respectively. 
Our results support this re-classification, and thus these two sources should be considered from now onward as 
part of the classical SgXBs. The cumulative luminosity distributions of all classical SgXBs in Fig.~\ref{fig:cumul} have been 
plotted with thicker dashed lines. 

\item IGR J18483-0311 and IGR J17354-3255 have similar distributions as classical SgXBs, but their overall profile and the knees  
appear to be shifted at lower luminosities ($\sim$10$^{35}$ erg~s$^{-1}$). These two sources are classified as  ``intermediate'' SFXTs, and 
are thus thought to be the missing link between the SFXTs and classical SgXBs \citep[due to their reduced 
dynamic range in the X-ray luminosity; see, e.g.,][]{rahoui08b,giunta09,ducci13}. The cumulative luminosity distributions of 
these two sources have been plotted in Fig.~\ref{fig:cumul} by using dot-dashed lines.

\item A similar conclusion as above applies to the cumulative luminosity distributions of the SFXTs IGR J16479-4514, 
IGR J16328-4726, and AX J1841.0-0536. The reduction in the average luminosity of IGR J16479-4514 and IGR J16328-4726 is evident 
once a comparison is carried out with, e.g., IGR J16418-4532 in the present paper and Vela X-1 in P14, respectively 
(note that Vela X-1 as an orbital period close to IGR J16328-4726, while IGR J16418-4532 has a period similar to IGR J16479-4514). 
No orbital period is known yet in the case of AX J1841.0-0536. The cumulative luminosity distributions of these SFXTs 
are plotted in Fig.~\ref{fig:cumul} with dotted lines. 

\item the cumulative luminosity distributions of the SFXT prototypes IGR J17544-2619, XTE J1739-302, and IGR J08408-4503 are shifted 
to even lower luminosities than other sources in this class. These three objects also display somewhat more complex profiles, and the 
identification of a knee is not trivial as in all other cases. We used solid lines in Fig.~\ref{fig:cumul} to represent the luminosity 
distributions of the three SFXT prototypes.     
\end{itemize}

Given the complex variety of all the cumulative distribution profiles, we did not attempt to fit them with some phenomenological 
model (e.g. a single or broken power-law). Instead, we show below how the shape of these profiles gives precious insights 
on the physical mechanisms regulating the X-ray activity of classical SgXBs and SFXTs.

\section{Discussion and conclusions}
\label{sec:discussion}

We made use of the long-term monitoring observations performed with the XRT on-board \swift\ to construct for the 
first time the cumulative luminosity distributions of most of the currently known SFXTs and a few classical SgXBs. 
Because of the re-classification of the sources IGR J16418-4532 and IGR J16465-4507, the cumulative luminosity 
distribution of three classical SgXBs can be presently constructed by using XRT data. The profile of these distributions 
closely resembles those of other classical SgXBs monitored in the hard X-rays and reported by P14 
(see, e.g, the cases of Vela X-1 and 4U 1700-377). 
This similarity suggests that the cumulative distributions of all classical SgXBs is generally characterized  
by a profile featuring a single-knee. The latter occurs at luminosities of $\sim$10$^{36}$-10$^{37}$~erg~s$^{-1}$. 

Single-knee profiles can be relatively well understood in terms of wind accretion from an inhomogeneous medium.  
\citet{furst10} showed that the X-ray luminosity of a system in which the NS is accreting from a highly 
structured medium, rather than a smooth wind, is expected to have a typical log-normal distribution. The profile of the corresponding 
cumulative distribution would thus be characterized by the presence of a single knee. Structures in the winds of a supergiant star are 
usually associated with ``clumps'', i.e. regions endowed with larger densities (a factor of $\sim$10) and different velocities (a factor of few) 
with respect to the surrounding medium \citep{ocr1988}. These structures can be as large as $\sim$0.1~R$_{*}$ 
\citep[here R$_{*}$ is the radius of the supergiant star; see, e.g.,][]{dessart02,dessart03,dessart05,surlan13}. 
According to the classical picture of wind accreting systems \citep[see, e.g.,][and references therein]{frank02}, the variation in the local 
density and/or velocity around a compact object produced by a clump can give rise to rapid changes in the mass accretion 
rate and thus on the released X-ray luminosity. Accretion from a moderately clumpy wind can thus qualitatively explain the X-ray variability 
of SgXBs and the profile of their cumulative luminosity distributions.  

\citet{oskinova12} showed that, despite the remarkable variations in the X-ray luminosity that can be produced by accretion 
from a highly inhomogeneous medium, the long-term averaged luminosity of the system is comparable to that obtained in the 
case of a smoothed-out wind. It is thus expected that the position of the knee in the cumulative luminosity distribution 
of a SgXB, being roughly associated to the value of its 
averaged X-ray luminosity, will mainly depend on its orbital period: the closer the NS to its companion, the higher the 
expected averaged X-ray luminosity\footnote{We neglected here the eccentricity, photo-ionization of X-rays on the supergiant 
wind and other processes that can affect the overall X-ray luminosity  
\citep[see, e.g.,][and references therein]{ducci10}. 
A detailed treatment of these effects is beyond the scope of the present paper.} (due to the enhanced density and slower 
velocity of the wind). This trend seems to be qualitatively respected by the classical SgXBs in our Fig.~\ref{fig:cumul} 
and in Fig.~1 of P14. As an example, the knee of IGR J16418-4532, which is characterized by an orbital period of 3.4 days, is located at a 
higher luminosity with respect to that of IGR J18027-2016, which has a larger orbital period (4.5~days). The same is true if the 
comparison is carried out between IGR J18027-2016 and Vela X-1 (orbital period 8.9~days), and if the even larger orbital period of 
IGR J16465-4507 is taken into account. Additional XRT monitoring observations of classical SgXBs are currently being planned in 
order to confirm these findings. 

According to the discussion above, it is unlikely that a simplified accretion wind scenario including only the presence of clumps 
could explain the X-ray behavior observed from the SFXT sources. Clumps provide, in principle, the means to trigger  
SFXT outbursts, but they cannot account for the substantial lower luminosity of these sources compared to classical SgXBs. 
To corroborate this argument, we first consider the cumulative distributions of the intermediate SFXTs IGR J18483-0311 and  
IGR J17354-3255, which are thought to be the missing link between classical systems and the SFXT prototypes. 
The profile of the distributions displayed by these two sources are similar to those of classical SgXBs, 
but are shifted toward the lower left side of the plots in Fig.~\ref{fig:cumul}.  
As an example, IGR J17354-3255 is characterized by an orbital period close to that of Vela X-1, but its   
average X-ray luminosity is a factor of $\sim$10 lower (see Fig.~1 in P14). 
This problem worsens when the cumulative luminosity distributions of the other 
SFXTs are considered. All SFXTs observed by XRT appear to be on-average much less luminous than the classical 
SgXBs. It is particularly worth mentioning the case of IGR J16479-4514 which has an orbital period similar to that of IGR J16418-4532 but its 
luminosity distribution is shifted at an average luminosity that is roughly a factor of $\sim$100 lower. 
The same conclusion would be reached by comparing the SFXT prototype 
IGR J17544-2619 with IGR J18027-2016 which have similar orbital periods (note that the relatively small 
uncertainties on the distance to all sources considered here would not be able to compensate for the estimated differences 
in luminosity; see Table~6 in R14). Beside being characterized by the lowest average luminosity, the three SFXT prototypes show also 
cumulative luminosity distributions with relatively complex profiles. In these cases, it is not trivial to accurately identify the 
main knee of their distribution. 

It is interesting to note that the distributions of all SFXTs in Fig.~\ref{fig:cumul} would clearly lead to low activity DCs for these objects 
when observed through low sensitivity large FoV instruments\footnote{The sensitivity limit is different for each source, as it depends 
on the intrinsic flux and the exposure time considered. We refer the reader to P14 for an exhaustive discussion regarding the ISGRI sensitivity 
limits for the observations of SFXTs.}. The latter are, indeed, not able to probe the rapid increases of the cumulative luminosity 
distributions of these sources in their fainter luminosity states, thus permitting us to study only 
the power-law shaped decay above $\gtrsim$10$^{35}$ erg~s$^{-1}$ (see P14). 

The mass loss rate of supergiants is known to have a significant spread depending on the star properties \citep{vink2000,puls08}.  
However, the fact that all SFXTs are characterized by similar companion stars to those in classical SgXBs \citep{rahoui08b} 
but are significantly sub-luminous compared to them, suggests a difference in 
the accretion processes on-going in these sources rather then a systematic discrepancy in the physical properties of their 
stellar winds (e.g., clumping factors). 
In order to produce a large decrease in the long-term X-ray luminosity, 
a mechanism is required to inhibit at least part of the accretion toward the NS and regulate plasma entry within the compact star magnetosphere. 
Theoretical models suggested so far to interpret the X-ray variability of SFXTs provide different ways to account for this feature. 

In the models proposed by \citet{grebenev07} and \citet{bozzo08b}, the inhibition of accretion is provided by the onset of 
centrifugal and/or magnetic barriers. 
The latter are due to the rotation and magnetic field of the NS. Depending on the strength of this field and the value of 
the spin period, the onset of different accretion regimes can lead to a substantial variation of the overall source luminosity 
(a factor of 10$^4$-10$^5$). The switch from one regime to another is triggered by the interaction of the NS with moderately 
dense clumps. Assuming typical parameters of supergiant star winds, the largest variability is achieved when the magnetic barrier 
is at work. The latter requires intense magnetic fields ($\gtrsim$10$^{14}$~G) and long spin periods ($\gtrsim$1000~s). 
While the magnetic gating would easily provide the means to achieve an X-ray variability comparable to that shown by the SFXT prototypes, 
the recent discovery of a cyclotron line at $\sim$17~keV from IGR\,J17544-2619 (suggesting a NS magnetic field intensity as low as  
B$\simeq$10$^{12}$~G) raised questions on the applicability of the magnetic gating model at least to this source \citep{bhalerao14}. 
  
In the quasi-spherical settling accretion model proposed by 
\citet{shakura12}, the inhibition of accretion is provided by a hot quasi-static shell that forms above the NS magnetosphere when 
a sufficiently low mass accretion rate is maintained.  
A substantial average reduction (a factor of $\sim$30) of the mass accretion rate onto the NS (and thus X-ray luminosity) is expected 
if the plasma entry through the compact star magnetosphere from the shell is regulated by inefficient radiative plasma cooling. 
If Compton cooling dominates, a reduction of the mass accretion rate by a factor of $\sim$3 is achieved \citep{shakura13}. 
The bright SFXT flares are proposed to result from sporadic reconnections between the NS 
magnetosphere and the magnetic field embedded in the stellar wind. According to this model, the main difference between SgXBs and 
SFXTs would thus be that only for the latter sources the wind properties are such that a low density is stably maintained around the 
compact object (e.g., through a systematically lower mass loss rate from the supergiant star or higher/lower wind velocity/density) 
and magnetized stellar winds play a role in triggering large accretion episodes \citep{shakura14}. However, such requirements are difficult 
to accommodate, given the lack of any clear evidence of systematic differences between stellar winds in SFXTs and classical SgXBs 
(see Sect.~\ref{sec:intro}). Further theoretical studies are currently on-going to investigate these issues.  

Finally we note that the cumulative luminosity distributions of the SFXT prototypes reported in Fig.~\ref{fig:cumul} 
feature the presence of plateau and multiple knees and thus look more complex than the profiles of other SFXTs and classical systems. 
At present we cannot exclude that these plateau are due to the relatively low number of bright SFXT outbursts 
recorded by XRT, which limits the completeness of the cumulative distributions  at the higher luminosities 
($\gtrsim$10$^{36}$ erg s$^{-1}$; see also R14). In case future outbursts detected by XRT during our monitoring 
campaigns will be discovered to span a relatively large range in luminosity at the peak (e.g., a factor of 10 or more 
in the same time bin considered here), the decay of the cumulative luminosity distributions could be 
significantly affected (this would not change the sub-luminosity problem discussed before). However, it is noteworthy 
that the $\sim$12~years monitoring campaigns carried out with the \rxte/PCA on several SFXTs also feature plateau. 
Although the plateau in the PCA data are less prominent than those observed by XRT, in both cases these features are due 
to the brightest SFXT outbursts which are detected as rare events and span a relatively limited range in luminosity \citep{smith12}. 
If consolidated by future XRT monitoring observations, this could be interpreted in terms of those peculiar source  
states discussed above during which the highest mass accretion rate is achieved. 

We conclude that the currently available XRT data provide support in favor of the general features of the  
theoretical models proposed so far to interpret the SFXT behavior, but do not allow yet to distinguish between them.  
A number of open questions remain to be investigated theoretically in the near future, including the requirement of strong magnetic fields  
for the applicability of the magnetic gating and the need for systematic differences in stellar wind parameters in the settling accretion 
model.

\section*{Acknowledgments}

PR acknowledges contract ASI-INAF I/004/11/0.  
LD thanks Deutsches Zentrum f\"ur Luft und Raumfahrt (Grant FKZ 50 OG 1301). 
The XRT data were obtained through ToO observations 
(2007-2012; contracts ASI-INAF I/088/06/0, ASI-INAF I/009/10/0) 
and through the contract ASI-INAF I/004/11/0 (2011-2013, PI P.\ Romano).

\end{document}